\documentclass[longauth]{aa}

\usepackage{graphicx}
\usepackage{graphics}
\usepackage{txfonts}
\usepackage{epsfig}
\usepackage{natbib}
\usepackage{hyperref}

\usepackage[]{color}

\begin{document}

\title{An image of the dust sublimation region in the nucleus of NGC~1068}

\author{GRAVITY Collaboration 
\and O.~Pfuhl\inst{1}
\and R.~Davies\inst{1}
\and J.~Dexter\inst{1} 
\and H.~Netzer\inst{8} 
\and S.~H\"onig\inst{5} 
\and D.~Lutz\inst{1} 
\and M.~Schartmann\inst{1}
\and E.~Sturm\inst{1} 
\and A.~Amorim\inst{19,21} 
\and W.~Brandner\inst{22}  
\and Y.~Cl\'enet\inst{2} 
\and P.~T.~de~Zeeuw\inst{1,17} 
\and A.~Eckart\inst{3,18} 
\and F.~Eisenhauer\inst{1} 
\and N.M.~F\"orster~Schreiber\inst{1} 
\and F.~Gao\inst{1} 
\and P.~J.~V.~Garcia\inst{15,20,21} 
\and R.~Genzel\inst{1,4} 
\and S.~Gillessen\inst{1} 
\and D.~Gratadour\inst{2} 
\and M.~Kishimoto\inst{6} 
\and S.~Lacour\inst{2,1} 
\and F.~Millour\inst{7}  
\and T.~Ott\inst{1} 
\and T.~Paumard\inst{2} 
\and K.~Perraut\inst{12} 
\and G.~Perrin\inst{2} 
\and B.~M.~Peterson\inst{9,10,11} 
\and P.~O.~Petrucci\inst{12} 
\and M.~A.~Prieto\inst{23}  
\and D.~Rouan\inst{2} 
\and J.~Shangguan\inst{1}
\and T.~Shimizu\inst{1} 
\and A.~Sternberg\inst{13,14} 
\and O.~Straub\inst{1} 
\and C.~Straubmeier\inst{3} 
\and L.~J.~Tacconi\inst{1} 
\and K.~R.~W.~Tristram\inst{15}  
\and P.~Vermot\inst{2} 
\and I.~Waisberg\inst{1} 
\and F.~Widmann\inst{1} 
\and J.~Woillez\inst{16}}

\institute{
Max Planck Institute for Extraterrestrial Physics(MPE), Giessenbachstr.1, 85748 Garching, Germany
\and LESIA, Observatoire de Paris, Universit\'e PSL, CNRS, Sorbonne Universit\'e, Univ. Paris Diderot, Sorbonne Paris Cit\'e, 5 place Jules Janssen, 92195 Meudon, France
\and I. Institute of Physics, University of Cologne, Z\"ulpicher Stra{\ss}e 77,50937 Cologne, Germany
\and Departments of Physics and Astronomy, Le Conte Hall, University of California, Berkeley, CA 94720, USA
\and Department of Physics and Astronomy, University of Southampton, Southampton, UK
\and Department of Physics, Kyoto Sangyo University, Kita-ku, Japan
\and Universit\'e C\^ote d'Azur, Observatoire de la C\^ote d'Azur, CNRS, Laboratoire Lagrange, Nice, France
\and School of Physics and Astronomy, Tel Aviv University, Tel Aviv 69978, Israel
\and Department of Astronomy, The Ohio State University, Columbus, OH, USA
\and Center for Cosmology and AstroParticle Physics, The Ohio State University, Columbus, OH, USA
\and Space Telescope Science Institute, Baltimore, MD, USA
\and Univ. Grenoble Alpes, CNRS, IPAG, 38000 Grenoble, France
\and School of Physics and Astronomy, Tel Aviv University, Tel Aviv 69978, Israel
\and Center for Computational Astrophysics, Flatiron Institute, 162 5th Ave., New York, NY 10010, USA
\and European Southern Observatory, Casilla 19001, Santiago 19, Chile
\and European Southern Observatory, Karl-Schwarzschild-Str. 2, 85748 Garching, Germany
\and Sterrewacht Leiden, Leiden University, Postbus 9513, 2300 RA Leiden, The Netherlands
\and Max Planck Institute for Radio Astronomy, Bonn, Germany
\and Universidade de Lisboa - Faculdade de Ci\^{e}ncias, Campo Grande, 1749-016 Lisboa, Portugal
\and Faculdade de Engenharia, Universidade do Porto, rua Dr. Roberto Frias, 4200-465 Porto, Portugal
\and CENTRA - Centro de Astrof\'isica e Gravita\c{c}\~{a}o, IST, Universidade de Lisboa, 1049-001 Lisboa, Portugal
\and Max Planck Institute for Astronomy, K\"onigstuhl 17, 69117, Heidelberg, Germany
\and Instituto de Astrof\'isica de Canarias (IAC), E-38200 La Laguna, Tenerife, Spain
}






\abstract{We present near-infrared interferometric data on the Seyfert 2 galaxy NGC~1068, obtained with the GRAVITY instrument on the European Southern Observatory Very Large Telescope Interferometer.
The extensive baseline coverage from 5 to 60\,$\rm M\lambda$ allowed us to reconstruct a continuum image of the nucleus with an unrivaled $0.2$\,pc resolution in the $K$-band. We find a thin ring-like structure of emission with a radius $r=0.24\pm0.03$\,pc, inclination $i=70\pm5^\circ$, position angle $PA=-50\pm4^\circ$, and $h/r<0.14$, which we associate with the dust sublimation region. The observed morphology is inconsistent with the expected signatures of a geometrically and optically thick torus. Instead, the infrared emission shows a striking resemblance to the 22\,GHz maser disc, which suggests they share a common region of origin. The near-infrared spectral energy distribution indicates a bolometric luminosity of $\rm (0.4-4.7) \times 10^{45}\,erg/s$, behind a large $A_K\approx5.5$ ($A_V\approx90$) screen of extinction that also appears to contribute significantly to obscuring the broad line region.} 

\titlerunning{An image of the dust sublimation region in the nucleus of NGC~1068}
\authorrunning{GRAVITY Collaboration}

\maketitle


%
\section{Introduction} 
\label{sec:intro}

NGC~1068 is often regarded as an archetypical Seyfert 2 galaxy \citep[e.g.][]{bland-hawthorn1997}. In particular, observations of NGC~1068 played an important role in the early development of unified models of active galactic nuclei (AGN).
In this model, opaque material in the equatorial plane obscures the supermassive black hole, accretion disc, and broad emission lines, so that the the AGN can only be observed directly from polar directions \citep{miller1983,antonucci1985}. Following the inception of the torus concept \citep{antonucci1982,antonucci1984} and the first efforts to explain its physical properties \citep{krolik1988}, there have been numerous observations of the central structures in this and other AGN from across the electromagnetic spectrum and covering many different tracers. 
Ideas about the geometry of the central obscurer have evolved and modified significantly over the years. Indeed, in the literature the term `torus' is used to mean a variety of different things. While it was always understood that this structure might be clumpy \citep{krolik1988}, early calculations adopted a smooth geometry \citep{pier1992}. The subsequent incorporation of a clumpy structure led to a suite of models that provide very good fits to the near-to-mid infrared spectral energy distribution (SED) of many AGN \citep{nenkova2008,honig2006,stalevski2012}, and which are also consistent with dust reverberation measurements \citep{koshida2014}.
We refer to \cite{netzer2015} for a detailed review.
Models and observations have achieved a level of detail where it is possible to apply a wide range of tests to the properties and assumptions used in creating them.
Additionally, the data we present here allow a new test to be applied. It concerns the observed morphology of the hot dust close around the AGN.
To provide a context for this, we highlight a few key observations pertinent to understanding the central few parsecs around the black hole in NGC~1068 as follows.

The detection of broad permitted emission lines in polarised light \citep{antonucci1985} represents a cornerstone of the AGN unification scheme \citep{antonucci1993}, which postulates that the observed properties of an AGN mainly depend on inclination. \cite{cameron1993} pointed out that other structures must also be present on small scales in NGC~1068 since the mid-infrared (MIR) continuum is extended on scales of $\sim1$\arcsec\ rather than dominated by a central source. Using higher resolution data, \cite{bock2000} showed there was a central $<0.2$\arcsec\ core, which they associated with the torus, and a tongue extending $\sim1$\arcsec\ along a north-south axis that mimics the inner part of the radio jet \citep{gallimore1996}, which is thought to be launched in a northerly direction and bends to the north-east when it encounters a dense molecular cloud.

Early K-band speckle and interferometry work found hot dust emission coming from multiple components on scales of 400\,mas to a few mas \citep{weigelt2004,wittkowski2004}. 
Observations with MIDI at 10\,$\mu$m indicated the presence of multiple dust components with a central 600-800\,K elongated dust core. 
The first component is a 800\,K, disc-like structure about $1.4 \times 0.5$\,pc in size  at a position angle (PA) of approximately $-45^{\circ}$ \citep{jaffe2004,raban2009,lopez-gonzaga2014}. A second, cooler $\sim300$\,K component, $3\times4$\,pc in size, was identified with the `body' of the torus \citep{raban2009}.

The presence of a complex structure consisting of multiple components is supported by X-ray spectra, indicating a compact high column density $N_{H}=10^{25}\,\rm cm^{-2}$ X-ray reflector with lower column $N_{H}=10^{23}\,\rm cm^{-2}$ clouds on more extended scales \citep{bauer2015}.
This does not necessarily imply anything about a dusty structure, since the X-ray absorption could be through dust-free clouds in the optically hidden broad line region (BLR).
The lack of Br$\alpha$ detection led \cite{lutz2000a} to put a lower limit on the extinction corresponding to $A_V \sim 100$\,mag, or an equivalent column of $n_H \sim 10^{23}$\,cm$^{-2}$ for a standard dust-to-gas ratio.

A maser disc is seen extending up to $\sim15$\,mas (1\,pc) from the AGN, with velocities reaching $\pm$330\,km\,s$^{-1}$ \citep{gallimore1996maser,greenhill1997,gallimore2004}. In addition, \cite{greenhill1997} noted that the $\sim-45^\circ$ position angle of the disc rotation axis differs from that of the jet by 30--$40^\circ$.
As such, there is clearly some warping on sub-pc scales in NGC~1068, and any description of the geometry needs to be associated with a particular structure or radial scale.
The maser disc can be described by a thin edge-on ($i \sim 80^\circ$) disc. Studies based on SED modelling also favour torus inclinations between $70^\circ$ \citep{hoenig2007,lopez2018} and $90^\circ$ \citep{hoenig2008}.
Attempts to model the three dimensional orientation of the system found that the northern narrow line region (NLR) cone is directed towards the observer and the southern part away from the observer \citep{packham1997,kishimoto1999}.

Near-infrared (NIR) integral field adaptive optics observations revealed streamers in the H$_2$ 1-0\,S(1) line, tracing hot molecular gas, which were interpreted as infalling material on a few tens of parsecs scales \citep{muller-sanchez2009}. 
They also provided the first glimpse of a molecular structure on a $\sim10$\,pc scale.

\cite{gratadour2015} obtained polarimetric imaging with SPHERE. This showed a bicone structure perpendicular to the maser disc, probably related to outflows, as well as an elongated patch aligned with the central molecular structure but extending further out ($R\approx25$\,pc), which they interpreted as the less dense outskirt of the putative torus.

ALMA observations of the $432\,\rm \mu m$ continuum and CO(6-5) molecular line resolved a 7--10\,pc thick-disc like structure \citep{gallimore2016,garcia-burillo2016}. As expected for a Seyfert 2 galaxy, observational data argue for a system close to edge-on.
Based on the linewidth of the spatially resolved HCN and HCO$^+$ lines in more recent observations of this structure, \cite{imanishi2018} argue for turbulent motions with $\sigma = 70$--80\,km\,s$^{-1}$, of similar order as the rotational velocity.
\citet{impellizzeri2019} find similar line widths close to the nucleus.
Garc\'ia-Burillo et al. (in prep.) detected the molecular disc in CO(2-1) and (3-2) at a PA $\sim$115\,$^\circ$ with an extension of 30\,pc.

Taken together, these observations, while still being consistent with and requiring the existence of a nuclear obscuring structure, are incompatible with geometrically thick clumpy torus models that attempt to account for all of the nuclear near-to-mid infrared continuum.
\cite{vollmer2018} propose a resolution to this discrepancy by constructing a model which has a thin disc on scales $\la 1\,\rm pc$ and a thick disc on scales of 1--10\,pc, the structure of which is determined by inflow from larger scales in the host galaxy. In their model, a magnetocentrifugal wind is launched at the boundary between the discs, and accounts for the elongated polar structures seen in many MIR interferometric measurements \citep{lopez-gonzaga2016}. This multi-component model was able to account for a wide variety of detailed observations when applied to NGC~1068.
A similar approach has been described by \cite{hoenig2019} who posits a dusty disc with small to moderate scale height, through which gas flows inwards and is then unbound in a dusty wind launched at the inner puffed-up rim by radiation pressure from the AGN.

\section{Prerequisites and observations}

\subsection{Bolometric luminosity and sublimation distance}
\label{sec:bol_lum}

For consistency with most of the NGC~1068 literature, we adopt the `standard' distance of 14.4\,Mpc \citep{bland-hawthorn1997}  but note that $d\approx16.5$\,Mpc would arise from Virgocentric infall velocity 1117\,km\,s$^{-1}$ \citep[][{\it NED}\footnote{The NASA/IPAC Extragalactic Database (NED) is operated by the Jet Propulsion Laboratory, California Institute of Technology, under contract with the National Aeronautics and Space Administration.}]{mould2000} and the Hubble constant ($H_0 = 67.8 \pm 0.9\, \rm km~s^{-1}Mpc^{-1}$) of \cite{planck2016}.

The estimated black hole mass of up to $M_{\rm BH}\approx 1.7 \times 10^7\,\rm M_{\odot}$ (see Table\,\ref{tab:tab1}) corresponds, for a solar metallicity gas,  to an Eddington luminosity of $L_{\rm Edd}\approx 2.5\times10^{45}$\,erg/s.

The literature reports bolometric luminosities between $L_{\rm bol}\approx0.5 \times 10^{45}$\, erg/s \citep{lopez2018} based on SED modelling and $L_{\rm bol}\approx1.6 \times 10^{45}$\, erg/s based on MIR data \citep{raban2009}. \cite{prieto2010} on the other hand found a significantly lower luminosity of $\sim9\times10^{43}$\,erg/s. 

Our estimates of $L_{\rm bol}$ are based on various X-ray and optical observations of the source. 
When we compared typical luminosity probes, we found that the X-ray based $L_{\rm bol}$ estimates are significantly lower than the estimates from ionised lines.
Based on detailed modelling, \cite{bauer2015} infer an X-ray luminosity of $L_{\rm 2-10\,keV}\approx2\times 10^{43}$\,erg/s, which includes a number of X-ray components and a detailed SED fitting of the source. 
Using broadband X-ray data \cite{ricci2017} find an intrinsic $L_{14-195\,keV} \sim 5\times10^{42}$\,erg\,s$^{-1}$.
Moreover, recent observations of \cite{marinucci2016} find X-ray variability, due to column density variations. The authors infer an intrinsic $L_{\rm 2-10\,keV}\approx7\times 10^{43}$\,erg/s. The X-ray estimates combined with the \cite{marconi2004} bolometric correction factor, give $L_{\rm bol}=\rm 0.4-1.4\times 10^{45}$\,erg/s. 
An estimate based on the nuclear luminosity of the 12\,$\mu$m continuum can be made using the relations of \cite{asmus2015}, together with additional relations from \cite{winter2012}. These yield $L_{\rm bol} \sim 1.8\times10^{45}$\,erg\,s$^{-1}$.
For the optical data we use estimates based on two combinations of three narrow emission lines: $\rm H\beta$ $\lambda4861$, [OIII] $\lambda5007$, [OI] $\lambda6300$. The expressions are taken from \cite{netzer2009}  and make use of reddening corrected line fluxes and galactic type extinction. The line fluxes are taken from \cite{storchi-bergmann1995}. The resulting bolometric luminosities are $2.3\times 10^{45}$\,erg/s based on the [OIII] and [OI] lines, and $4.7\times 10^{45}$\,erg/s based on the $\rm H_{\beta}$ and [OIII] lines.
An early estimate, using scattered light from an off-nuclear cloud that arguably sees the hidden nucleus, also favoured luminosities rather in excess of $10^{45}$\,erg/s \citep{miller1991}.
The large range of luminosity estimates $L_{\rm bol}=0.4 - 4.7 \times 10^{45}$\, erg/s reflects the extreme extinction of the source.  In any case, NGC~1068 is radiating close to the Eddington limit.

Theoretical and observational work argue for a strong correlation between the NIR size and the AGN luminosity (\citealt{suganuma2006,kishimoto2011,koshida2014,GRAVITY2019_size}). In the standard picture, the NIR emission originates from the dust sublimation region, which is determined by the dust sublimation temperature $T_{\rm sub}$. The temperature at which dust sublimates depends on the grain size and composition and ranges from $T_{\rm sub} \approx1500$\,K for silicate (Si) grains to $T_{\rm sub} \approx2000$\,K for graphite (C) grains \citep{baskin2018}. By assuming a grain size distribution typical for the interstellar medium (ISM), the mean sublimation distance can be calculated for graphite and silicate dust as $R\approx a_{X}\times L_{46}^{1/2}$\,[pc] with $a_{\rm C} =0.5$ and $a_{\rm Si}=1.3$ (e.g. \citealt{barvainis1987,netzer2015}; for grain size dependence see \citealt{baskin2018}). 
However, there is evidence both from SED fitting \citep{mor2012} and from dust reverberation mapping \citep{kishimoto2007, koshida2014} that, close to the AGN, one may find only larger graphite grains.
Since, at the same distance from the AGN, large grains are cooler than small grains, one might naturally expect to find a distribution of large graphite grains at the smallest radii.
In addition changes in the brightness of the AGN may lead to changes in the dust sublimation radius \citep{kishimoto2013} or instead, depending on the dust distribution, its temperature \citep{schnuelle2015}.
As such, the sublimation radius is really rather a sublimation region according to grain size and species \citep[e.g.][]{hoenig2017,baskin2018}.
Empirical studies have found $R_{\rm 2.2 \mu m}\approx 0.4\,{\rm pc}\times (L_{\rm AGN}/10^{46} {\rm erg/s})^{1/2}$ (assuming the V-band to $R_{\rm sub}$ relation from \citealt{koshida2014}, and using the conversion $L_{\rm AGN}\sim 8L_{\rm 5500\AA}$ from \citealt{netzer2015}), which corresponds to a good approximation to the sublimation radius of graphite.
For NGC~1068 the observed relation from \cite{koshida2014} predicts a dust sublimation radius in the range $R_{\rm 2.2 \mu m}\approx 0.08 - 0.27$\,pc, for the luminosity range quoted in Table\,\ref{tab:tab1}.

\begin{table}
\begin{tabular}{lll}
\hline
\cline{1-2}
Parameter & Value &  Source \\
\hline
Distance     & $14.4$\,Mpc                         & [1]\\
$M_{\rm BH}$  & $ 0.8-1.7 \times 10^{7}\, M_{\odot}$ & [2], [6], [7], [8] \\
$L_{\rm bol}$ & $0.4- 4.7 \times10^{45}$\,erg/s     & Sec.~\ref{sec:bol_lum}, [3], [4], [5] \\
\hline
\end{tabular}
\caption{Basic parameters of NGC~1068 taken from the literature. References are 
[1]: \cite{bland-hawthorn1997}; 
[2]: \cite{lodato2003}; 
[3]: \cite{marconi2004}; 
[4]: \cite{bauer2015}; 
[5]: \cite{lopez2018}; 
[6]: \cite{greenhill1996}; 
[7]: \cite{hure2002}; 
[8]: Gallimore \& Impellizzeri (submitted).}
\end{table}\label{tab:tab1}

\subsection{Measured column densities}
\label{sec:column}

Numerous studies have tried to estimate the column density obscuring the line-of-sight to the central region of NGC~1068.
Early CO/HCN measurements of the central few arcsec suggested column densities of $N_H\sim 0.2-1\times 10^{23}\, \rm cm^{-2}$ \citep{planesas1991,tacconi1994,sternberg1994}. Because the beam is relatively large, these measurements may be affected by the complex structures in the nuclear region, for example the bright CO emission about 1\arcsec\ west of the nucleus. More recent higher resolution observations focus on the central molecular structure. \cite{imanishi2018} found gas masses of 
$2 \times 10^6\,M_\odot$ within 14\,pc based on ALMA observations of HCN 3-2 and HCO tracers, which imply a column density of 
$N_H \sim 2 \times 10^{23}\,\rm cm^{-2}$. 
ALMA $\rm 432\,\mu m$ continuum observations by \cite{garcia-burillo2016} inferred gas masses of 
$M_{\rm gas}=10^5\,M_{\odot}$ within a structure of 7\,pc diameter, suggesting a column density of 
$8 \times 10^{22}\,\rm cm^{-2}$. 

In the infrared, ISO upper limits on the Brackett $\alpha$ broad-line emission led \cite{lutz2000a} to place a lower limit on the extinction $A_{\rm Br\alpha,4.05}\geq2.4$, which corresponds to $A_{\rm K}\geq6$.   For an ISM-like gas-to-dust ratio, the optical extinction can be converted into a column density. Based on X-ray calibrations, the relationship between the hydrogen column density $N_H$ and the optical extinction $A_V$ is $N_H ({\rm cm}^{-2}) = a \times 10^{21} A_{\rm V} \rm \,(mag)$, where $a$ has a value ranging from about 1.79 to 2.21 \citep{predehl1995,guver2009}. Therefore, the observed extinction implies a hydrogen column density of 
$N_H \ga (1.8 \pm0.6) \times 10^{23}\, \rm cm^{-2}$.

The gas masses found by \cite{muller-sanchez2009} imply a column density of
$N_H \sim 3 \times 10^{24}\,\rm cm^{-2}$ assuming a source size of 0.2\arcsec\ and a hot-to-cold gas mass ratio of 
$M_{H_2}^{1-0S(1)}/M_{H_2}^{total} \sim 10^{-6}$. 
\cite{lopez2018} estimated 
$N_H = 5 \times 10^{23}\,\rm cm^{-2}$ based on 2--400$\rm\,\mu m$ clumpy torus SED fitting over a 0.4\arcsec\ aperture.

Generally larger column densities are found in the X-rays, which could partially originate from dust-free gas. Various studies found consistently a column density of 
$N_H \sim 1 \times 10^{25}\,\rm cm^{-2}$ (with a 50\% coverage) based on 3--100\,keV data \citep{matt1997,bauer2015} as well as based on broad-band 0.3--150\,keV X-ray model fitting \citep{ricci2017}. 

Overall the different measurements argue for a mean column density of $N_H \sim 10^{23}-10^{24}$\,cm$^{-2}$ within the central few parsecs to tens of parsecs.
The higher column density towards the X-ray emitting corona in the very centre may largely originate in dust-free gas \citep{risaliti02,netzer2015,davies15}.

\subsection{Extinction correction}
\label{sec:extinction}

Throughout the paper we have assumed the same extinction law as can be found in the highly absorbed Galactic Centre region \citep{fritz2011}. For wavelength $<2.6\,\rm\mu m$ we assumed an extinction slope of $A(\lambda)\propto \lambda^{-2.05}$. At larger wavelengths the extinction shows spectral features, which we interpolated based on the extinction published by \cite{fritz2011}. Some useful conversions are: $A_{\rm V}/A_{\rm Br \gamma}\approx20$, $A_{\rm Br \gamma}/A_{\rm Br \alpha}=2.5$, $A_{\rm Br \gamma}/A_{\rm 8.76\mu m}=1.22$ and $A_{\rm Br \gamma}/A_{\rm 12.4\mu m}=1.86$.

\subsection{Interferometric observations}
\label{sec:obs}

The observations were carried out with the GRAVITY instrument, which interferometrically combines either four 8-m Unit Telescopes (UTs), or four 1.8-m Auxiliary Telescopes (ATs), of the European Southern Observatory (ESO) Very Large Telescope (VLT). Observations of NGC~1068 included both of these, and provided 3~milli-arcsec (mas) resolution K-band (2.2\,$\rm \mu m$) continuum imaging. For a detailed description of the instrument and the data analysis we refer to \cite{GRAVITY2017}. In brief, the light of the four telescopes is extracted into mono-mode fibres \citep{pfuhl2012,pfuhl2014} for two positions on the sky and then interfered in the beam combiner for all six baselines of the interferometer.
The fibre diameters of 56\,mas and 250\,mas are matched to the 2.2\,$\rm \mu m$ diffraction limited beams of an 8-m UT and 1.8-m AT respectively. In a `normal' single-field observation, the light of the science target is split 50/50 and fed into two channels, the fringe-tracker and the science spectrometer. The fringe-tracker is used to correct the fast atmospheric optical path variations with typical integration times of 1-10\,milli-seconds \citep{lacour2019}. This allows coherent integration times with the science spectrometer up to $\sim$ 100\,s. The interferometric observations of NGC~1068 are challenging due to the complex nature of the object. The central 1\arcsec\ shows extended continuum as well as line emission. The bright core has been partially resolved with speckle data obtained at the 6-m telescope of the Special Astrophysical Observatory in Russia \citep{weigelt2004} as well as with adaptive optics assisted 8-m telescope data from the VLT \citep[e.g.][]{rouan2019}.  Overall there is a significant contribution of coherent flux on all spatial scales, which can be probed with the VLT interferometer (VLTI). The visibilities range from 44\% on the shortest AT baselines to $\sim$0.1\% on the longest UT baselines. In order to reach sufficient signal-to-noise, we decided to feed all the light to the fringe-tracker channel. This provided us with a factor two more signal on the fringe-tracker at the expense of spectral resolving power (The fringe-tracker has five spectral channels across the K-band, corresponding to $R\sim20$). 

The observations\footnote{ESO Telescopes at the La Silla Paranal Observatory programme IDs
0102.B-0667, 0102.C-0205 and 0102.C-0211.} were taken on 20 November 2018 (1:00-5:20 UTC) with the UT array and on 23 December 2018 (1:30-3:30 UTC) with the AT compact array. The UT observations on November 20th had excellent conditions with seeing measured by the differential image motion monitor (DIMM) between 0.47 - 0.65\arcsec\ and 7 - 13\,ms (optical) coherence time.  The conditions during the AT observations were almost as good with the seeing ranging from 0.43 - 0.75\arcsec\ and a coherence time between 4 - 8\,ms. In total we obtained 45\,min with the UTs and 30\,min with the ATs of integration time on source. The science observations were bracketed by observations of close calibrator stars HIP\,54 ($m_K=7.95$), HI\,P16739 ($m_K=8.93$) and HIP\,17272 ($m_K=8.34$). 
We used the standard pipeline \citep{2014SPIE.9146E..2DL,GRAVITY2017} to reduce and calibrate the data.

\subsection{Adaptive optics based photometry}
\label{sec:phot}

The combination of four telescopes allows us to simultaneously obtain the coherent as well as the incoherent (absolute) flux of the object. At the same time, the acquisition camera \citep{anugu2018} provides an estimate of the Strehl ratio (in H-band). By comparing the observed flux of NGC~1068 with observations of known calibrator stars HIP\,17272 ($m_K=8.34$), HIP\,16739 ($m_K=8.93$) and HIP\,54 ($m_K=7.95$) we derived an effective magnitude of NGC~1068. 
As previously described, NGC~1068 is resolved even on scales of a single telescope. This means that the fibre-injected flux depends on the size of the fibre mode on sky. With the UTs, we find a total flux of $310\pm80$\,mJy within the fibre beam ($D_{\rm fibre}\approx56\,\rm mas$). This corresponds to a magnitude of  $m_K=8.33\pm0.25$. Overall, our flux measurement is in good agreement with the previously reported flux of the central component 350\,mJy by \cite{weigelt2004} and with \cite{rouan2019}, who found a central $m_K=8.65$. 

\cite{weigelt2004} used speckle data to derive the flux distribution of the central few 100\,mas. They modelled the data with two elongated Gaussian components, a central component  with a full width at half maximum (FWHM) of $\rm 18/39\,mas$ contributing $F_1=350\rm \,mJy$ at $\rm PA\approx -20^\circ$, and a larger structure with a FWHM of $\rm 400\,mas$ that had a quite uncertain flux estimate of $F_2=30-230\rm \,mJy$.  The first component has a FWHM significantly smaller than the fibre mode (56/250\,mas). Therefore the component is fully injected into our GRAVITY fibres, which agrees with our flux measurement. The 400\,mas component however is only partially injected. Less than 30\% (9-69\,mJy) of the large component is injected into the fibres using ATs and only $\sim2$\% ($<$5\,mJy) is injected into the fibres using the UTs.

\section{Interferometric data and image reconstruction}
\label{sec:data}

\begin{figure*}[ht]
\includegraphics[width=\textwidth]{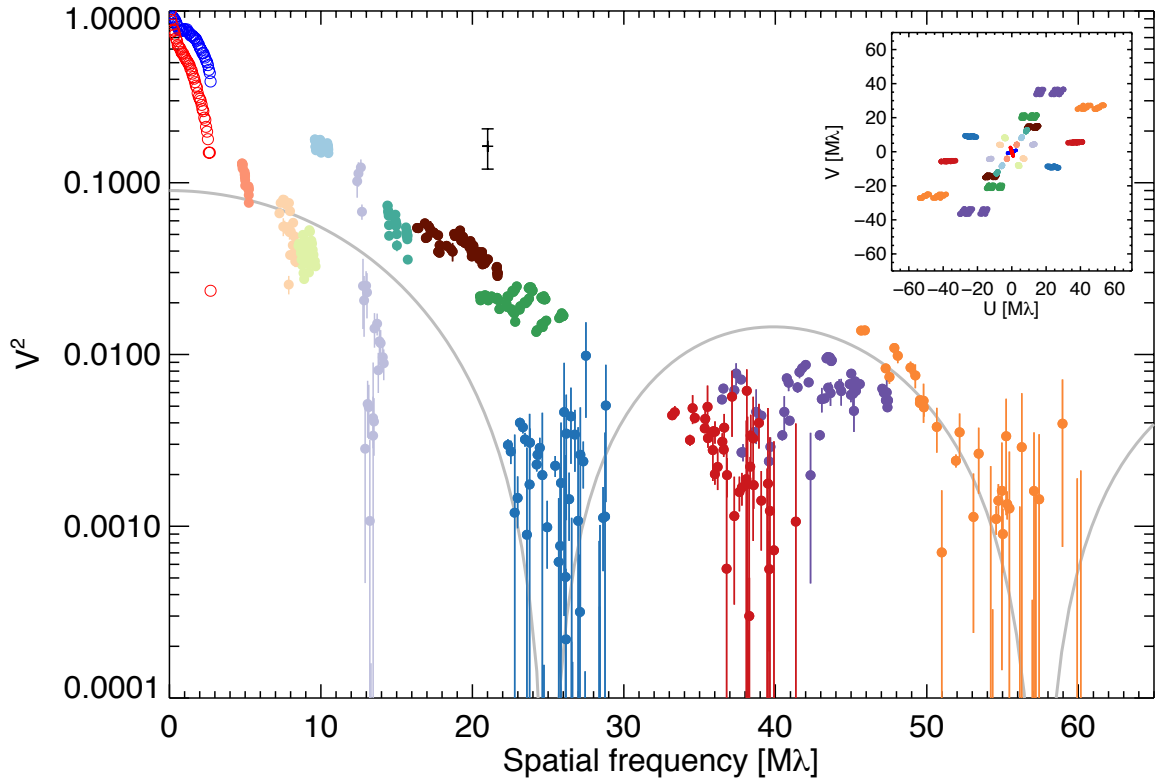}
\vspace{-2cm}
\caption{Squared visibilities measured for NGC\,1068.  The light coloured filled data points were obtained with the AT array in compact configuration (spatial frequencies $\rm 5<f<17\,M\lambda$.). The high spatial frequency data were taken with the UT array. The small inset shows the corresponding UV coordinates. The open blue and red symbols (spatial frequencies $\rm f<4\,M\lambda$) represent speckle data from \cite{weigelt2004}, which are shown only for comparison; they were not used for the image reconstruction. The data point and error bar at a spatial frequency of 21\,M$\lambda$, lying above our new data, shows the only previous near-infrared VLTI detection of NGC~1068 \citep{wittkowski2004}.
For comparison, the grey continuous line shows the visibility of a thin ring-shaped emission region with a radius of 3.3\,mas and a width of 0.5\,mas scaled to a maximum visibility of 0.3. While this reflects the general shape of the visibility distribution, it is a poor fit and indicates that more complex structures are present. 
\label{fig:Fig1}}
\end{figure*}
\begin{figure}[ht]
\hspace{-0.5cm}
\includegraphics[width=10cm]{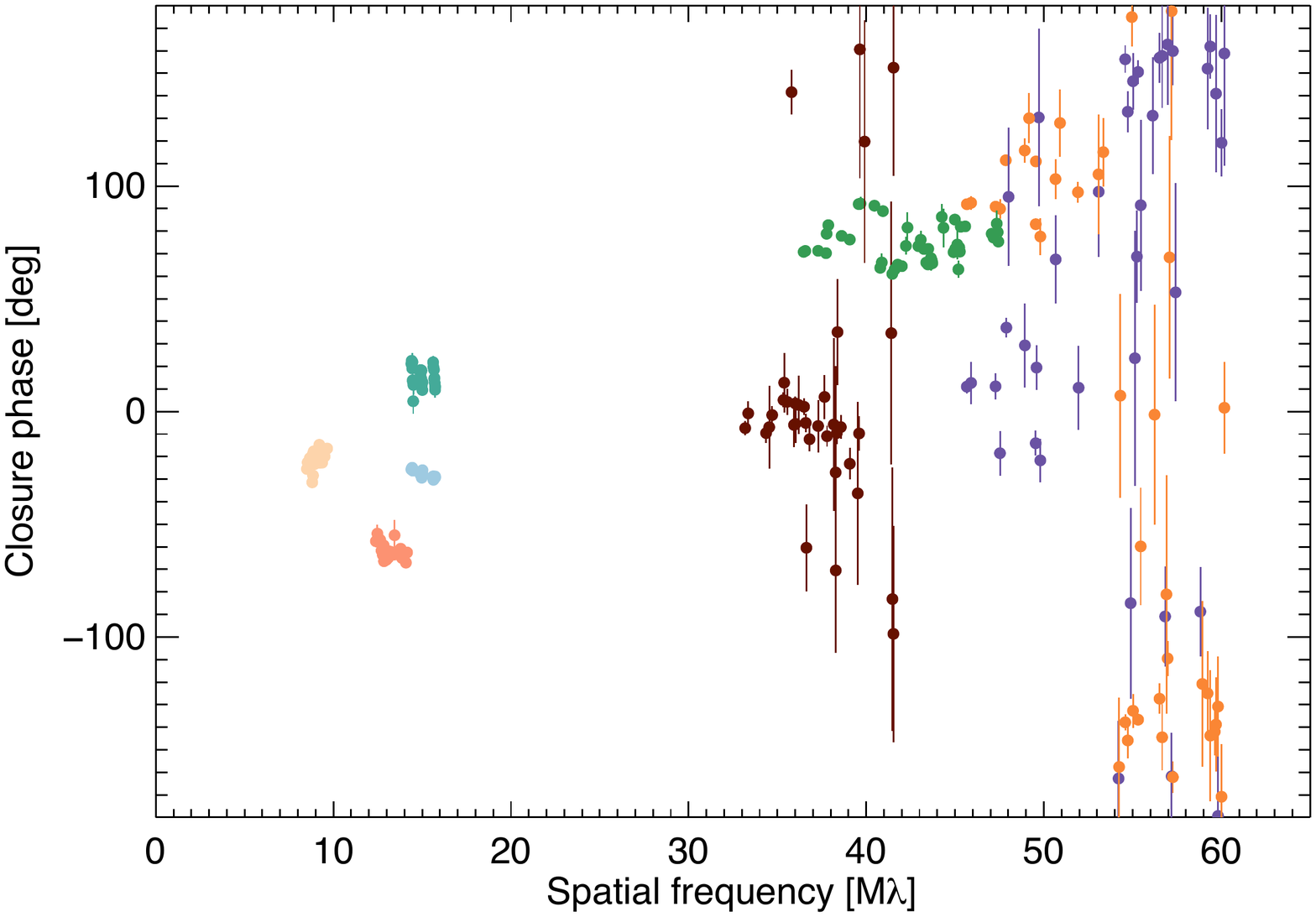}
\vspace{-1.5cm}
\caption{Closure phase data of NGC~1068. See Figure\,\ref{fig:Fig1} for the explanations of the symbols. The highly non-zero closure phases indicate strong source asymmetries on all spatial scales.}
\label{fig:Fig2}
\end{figure}

The interferometric data reveal significant structures on all spatial scales. In Figure\,\ref{fig:Fig1} we show the calibrated visibilities from the AT and UT observations. For comparison we show the early speckle data from \cite{weigelt2004} obtained with a single telescope. 
The observed visibilities cannot be approximated with a simple symmetric model like a Gaussian or a uniform disc. Also, as shown in the figure, a ring model fails to reproduce the data. This is however not surprising, given the large asymmetries in the source intensity distribution indicated by the highly non-zero closure phases (see Figure\,\ref{fig:Fig2}).
The complexity of the visibility and closure phase data prevent us from using simple model fitting to interpret the observations. Encouraged by the excellent quality of the data and the large uv coverage, we therefore turned to standard image reconstruction codes to analyse the data.

\subsection{An image of the dust sublimation region} 
\label{sec:img}

\begin{figure*}[ht]
\vspace{-0.5cm}
\includegraphics[width=\textwidth]{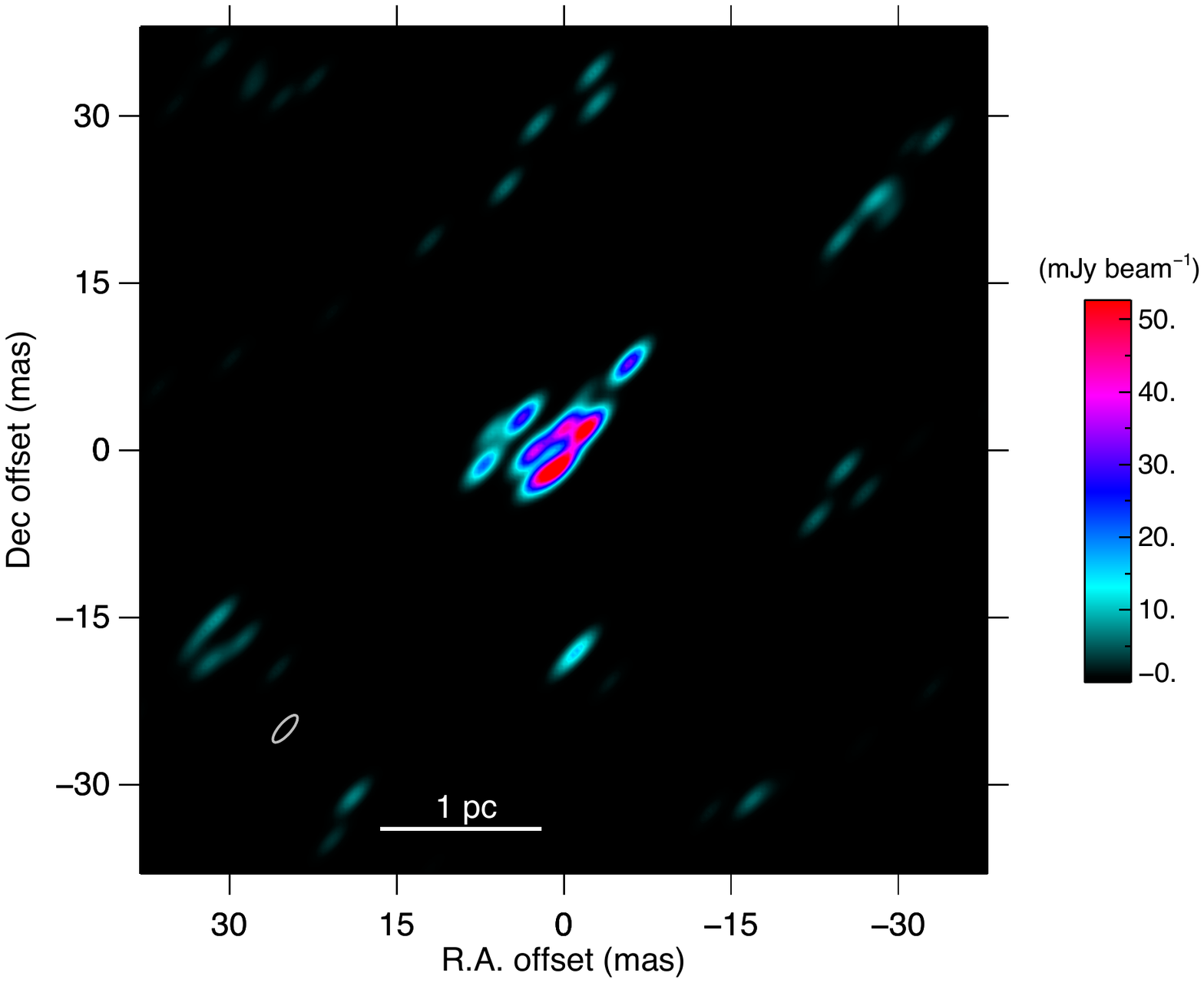}
\vspace{-2.5cm}
\caption{Reconstructed image of the central $5.1\times5.1$\,pc of NGC~1068. The elliptical beam, which was used to convolve the image, is shown in the lower left. Details of the reconstruction, which only allows positive sources, are given in Sec.~\ref{sec:img}. The central structures are robust to changing the reconstruction parameters. The fainter features towards the edge of the field, with fluxes close to the noise floor of $\sim 7$\,mJy\,beam$^{-1}$ RMS (i.e. dark cyan in this colour scale), are uncertain.}
\label{fig:Fig3}
\end{figure*}

\begin{figure*}[ht]
\vspace{-0.5cm}
\includegraphics[width=\textwidth]{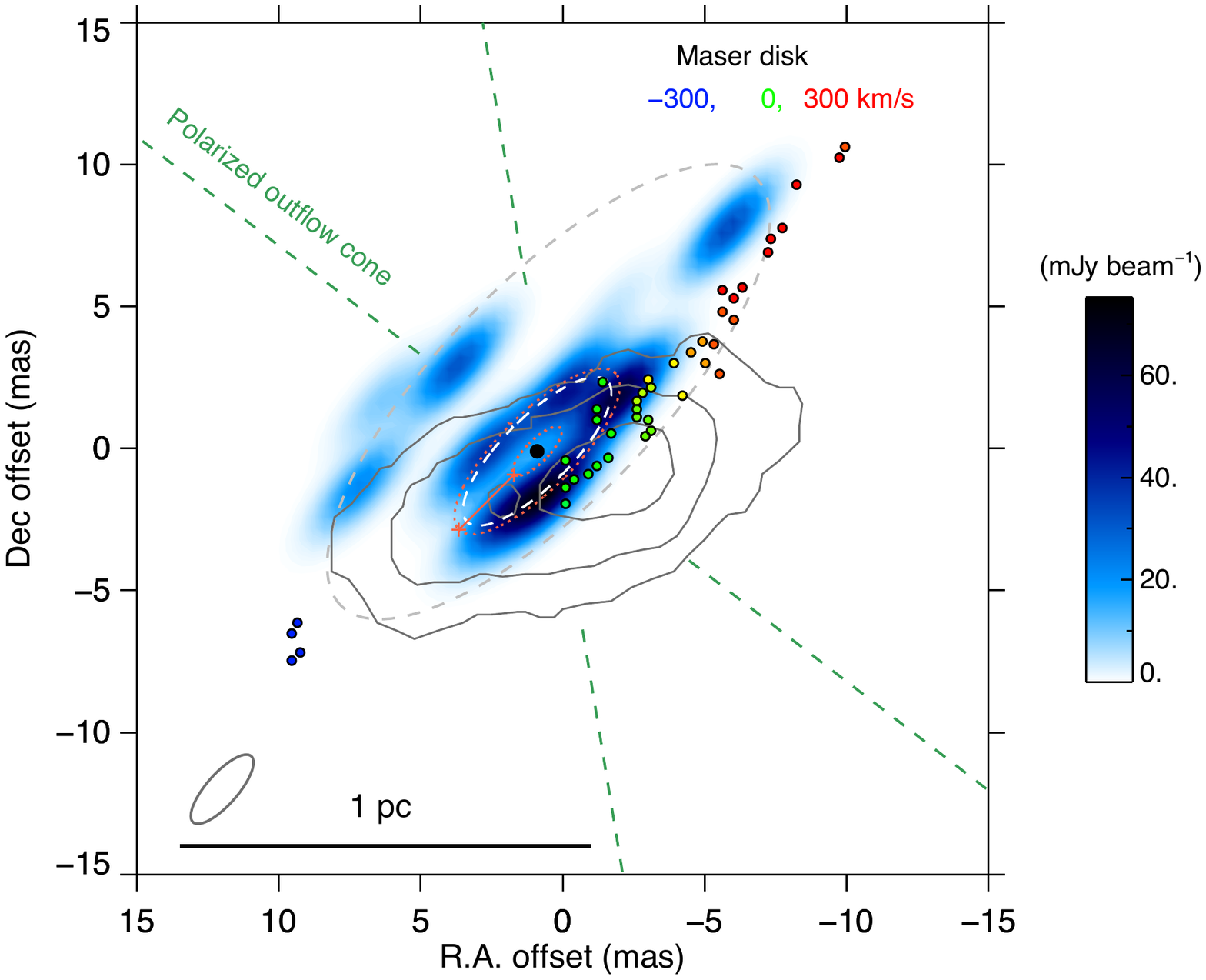}
\vspace{-2.5cm}
\caption{Reconstructed image of the inner $2.1\times2.1$\,pc of NGC~1068 (i.e. the central region of Fig.~\ref{fig:Fig3}). The best fitting circular ring is shown in dashed white ($r=0.24$\,pc (3.5\,mas).  The dust sublimation sizes for the range of bolometric luminosities ($\sim$0.08 - 0.27\,pc) are shown in orange dotted ellipses ($L_{\rm bol}=0.4 -4.7 \times 10^{45}\, \rm erg/s$). The filled black circle indicates the position of the AGN, corresponding to the kinematic centre of the masers . The radio continuum (grey contours) and the maser emission (filled coloured circles) were extracted from \cite{gallimore2004} and aligned based on the re-analysis of Gallimore \& Impellizzeri (submitted). The radio and IR images are aligned based on the continuum peaks. The green dashed lines indicate the bipolar outflow cone observed in IR polarisation on scales of few hundred mas \citep{gratadour2015}. The orientation of the polarisation cone is very similar to the large scale ionisation cone found by \citet{das2006}. The grey dashed ellipse indicates the observed $10\,\rm\mu m$ emission size \citep[taken from][]{raban2009}. The elliptical beam, used for the reconstruction, is shown in the lower left corner.}
\label{fig:Fig4}
\end{figure*}

We used the publicly available MiRA tool \citep{thiebaut2008} to reconstruct an image based on the visibility and closure phase data. A `grey image' was reconstructed assuming that the intensity distribution on sky $I(\alpha,\beta)$ is independent of the spectral channel across the observed K-band. We normalised the visibility by a factor 0.5 to match the visibilities at the smallest spatial frequencies. This means that we reconstruct 50\% of the photometric flux in our image ($\approx155$\,mJy).
The remaining flux originates at larger scales and/or is smoothly distributed. Figure\,\ref{fig:Fig3} shows the reconstructed image using MiRA with a {\it l2l1 smoothness} prior (threshold $5\times10^{-4}$) on a $200\times200$ pixel grid with a spatial scale of 0.4\,mas/pixel. As an initial guess, we used a single Gaussian image with 5\,mas FWHM. 
The final reconstructed image was then convolved with a Gaussian beam of FWHM $3.1 \times 1.1$\,mas and $\rm PA = 48.6^\circ$, which corresponds to 0.8 times the (UT array) interferometric beam.
In order to asses the systematic uncertainty of the image, we repeated the reconstruction using different penalty functions (for more information, refer to \citealt{thiebaut2008,thiebaut2017}) such as a {\it smoothness} prior, (i.e. a penalty on sharp transitions in the image), a {\it compact} prior (i.e. penalty on large radii) and a {\it l2l1 smoothness} prior (i.e. a prior able to maintain sharp transitions in the image). We stepwise optimised the gain and hyper parameters, where applicable. The resulting images varied but the central structure hardly changed in shape and amplitude.
In a second step, we determined the reliability of individual detections in the image by repeating the reconstruction with temporal and bandwidth restricted subsets of the data (with and without the bluest 1.995\,$\rm \mu m$ channel which is partially contaminated by metrology laser light). Spurious sources, which were not present throughout the repeated imaging attempts likely correspond to the reconstruction of noise in the data. This noise floor has a root-mean-square (RMS) of $\sim 7\,\rm mJy~beam^{-1}$. In addition, the image reconstruction methods used here, where calibrated phases are not available, assume positive fluxes. Therefore, a reconstructed image never contains negative values, unlike single-dish dark- or sky-subtracted images. Numerous spurious noise sources can be seen in Fig\,\ref{fig:Fig3}. The central ring-like structure is highly significant at $5-7\,\sigma$. The surrounding features to the north-east and north-west are significant at $3-4\,\sigma$. Figure\,\ref{fig:Fig4} represents a zoom on the significant central structure.

The image shown in Figure\,\ref{fig:Fig4} features an extended structure with a central hole, which can be described as a 3/4 ring with a radius of $r\approx 0.24\pm0.03$\,pc ($3.5\pm0.3$\,mas), position angle $PA=-50\pm4^{\circ}$ and an inclination of $i=70\pm5^{\circ}$. These parameters are summarised in Table~\ref{tab:ringpars}. The extent of the NIR structure resembles in size and orientation the observed $10\,\rm \mu m$ emission, yet at much higher resolution. The size of the inner bright ring-like structure matches remarkably well the expected dust sublimation distance of graphite grains in the radiation field of a central engine. The south-western ridge is about a factor two brighter than the north-eastern side of the ring-structure. The north-east orientation of the jet and the ionised outflow, which are approaching the observer \citep{gallimore2004}, suggests that the south-western ridge corresponds to the near-side of the emission ring (but see Sec.~\ref{sec:model4} for an alternative interpretation). The three dimensional orientation is reinforced if the radio continuum emission \citep{gallimore2004} is aligned close to the NIR emission. We find a striking resemblance between the maser emission locations and the bright south western rim of the K-band image. Since maser emission is beamed towards the observer, the emitting clouds must lie between the source (AGN) and the observer, and hence the maser emission should originate from the near-side.       
This alignment, discussed in Sec.~\ref{sec:maser}, supports the association of the ring-like structure with the dust sublimation region around the AGN, and it means that the near-side of the dusty ring is not only observable but also brighter than the far-side.
Those findings are very difficult to reconcile with geometrically thick clumpy torus models, which can only reproduce a NIR ring-like structure in systems that are relatively face-on, and struggle to make the near side of the ring brighter than the far side (the near and far sides are given by other observations, notably \cite{kishimoto1999} and \cite{gallimore2016}, as indicated in Sec.~\ref{sec:intro}).

\begin{table}
\begin{tabular}{ll}
\hline
\cline{1-2}
Parameter  & Value  \\
\hline
Radius ($r$)            & $0.24\pm 0.03$\,pc   \\
Thickness ($\delta r$)  & $ <0.035$\,pc        \\
Position angle $(PA)$ & $-50^\circ\pm4^\circ$ \\
Inclination $(i)$       & $70^\circ\pm5 ^\circ$ \\
\hline
\end{tabular}
\caption{Parameters of dust sublimation ring}
\label{tab:ringpars}
\end{table}


\subsection{Disc height}
\label{sec:discheight}

The observed width of the ring structure is consistent with it being unresolved. We can place a firm upper limit on the width of the ring of $<0.5$\,mas or 0.035\,pc.
If this is interpreted in terms of a vertical thickness, then the scale height of the ring must be $h/r<0.14$. 

\subsection{Cospatial dust and maser disc}
\label{sec:maser}

Maser emission requires a high velocity coherence of the maser emitting gas as well as a population inversion of the masering molecules \citep{lo2005}. 
 Energy and momentum conservation imply that the induced photon has the same frequency and direction as the stimulating photon \citep{elitzur1990,elitzur2002}.
There is a broad consensus that both the maser emission and the NIR emission are powered by the central AGN \citep[e.g.][]{neufeld1991}.
The X-rays are able to heat the gas above the 400\,K threshold where the water abundance is enhanced. To achieve a state in which atomic and molecular phases can co-exist, it is important that the medium is optically thin to the non-masing far-infrared photons, otherwise the trapping of these photons by the maser molecules would lead to a thermalisation of the population. Ultimately, the maser emission rate is limited by the loss or destruction
rate of this accompanying far-infrared radiation \citep{rank1971,neufeld1994}. This constrains the geometry of the maser source because the surface area
for cooling should be optimised, favouring an elongated or flat structure \citep{lo2005}.

The fact that mega-masers are almost exclusively found in Seyfert 2 galaxies \citep{ramolla2011} is consistent with the picture that disc maser emission is produced and beamed close to the plane of the obscuring material.
In NGC~1068, nuclear $\rm H_{2}O$ 22\,GHz maser emission is detected, spanning a velocity range of 600\,km\,s$^{-1}$ nearly symmetric about the systemic velocity of the galaxy \citep{greenhill1997,gallimore2001}. The maser spots extend out to $\sim$0.8\,pc, aligned along a position angle of roughly $-45^\circ$. While symmetric in velocity, the `red masers' at positive velocities are a factor $\sim$4-5 brighter than the `blue masers' \citep{gallimore2001}.
In particular $\rm H_{2}O$, 22\,GHz emission requires hydrogen densities of $\sim10^{7} - 10^{11}\,\rm cm^{-3}$ and dust temperatures above $\sim400$\,K, but not too far in excess of 1000-1500\,K \citep{neufeld2000}. The pump rate of different water emission lines (22, 183, 321 and 325\,GHz) depends sensitively on the local density and temperature \citep{neufeld2000}. The detection of these higher transitions in the future (e.g. with ALMA) can therefore provide valuable information on the conditions close to the sublimation region.

Unfortunately the absolute astrometric accuracy of the infrared emission (discussed in Sec.~\ref{sec:astrometry}) is not sufficient to align the radio and the infrared emission on the scales of interest for this paper, although the radio continuum and the maser emission can be aligned to $\sim$1.3\,$\rm mas$ accuracy \citep{gallimore2004}.
Given the lack of precise positioning information, one option for the relative alignment would be, given their striking similarities, to match the bright side of the ring-like structure in the K-band continuum with the high density of masers spots. However, this does not take into account the implied location of the central black hole. If one does so, a more natural choice is to align the centre of the ring with the kinematic centre of the masers (Gallimore \& Impellizzeri, submitted). The overall morphological match between the maser spots, the radio-continuum, and the south-western ridge (side facing the observer) of the K-band continuum is remarkable (see Figure\,\ref{fig:Fig4}). 
This suggests that the maser and the K-band emission originate from the same region, the dust sublimation ring. The fact that maser emission is only detected on the side facing the observer is consistent with the directional beaming of the maser emission (masing clouds are located between the central source and observer).
We only find infrared counterparts of the systemic masers and red masers. The weaker blue masers show no infrared counterpart.

\section{Infrared spectral energy distribution}

\begin{figure*}[ht]
\vspace{-0.5cm}
\includegraphics[width=\textwidth]{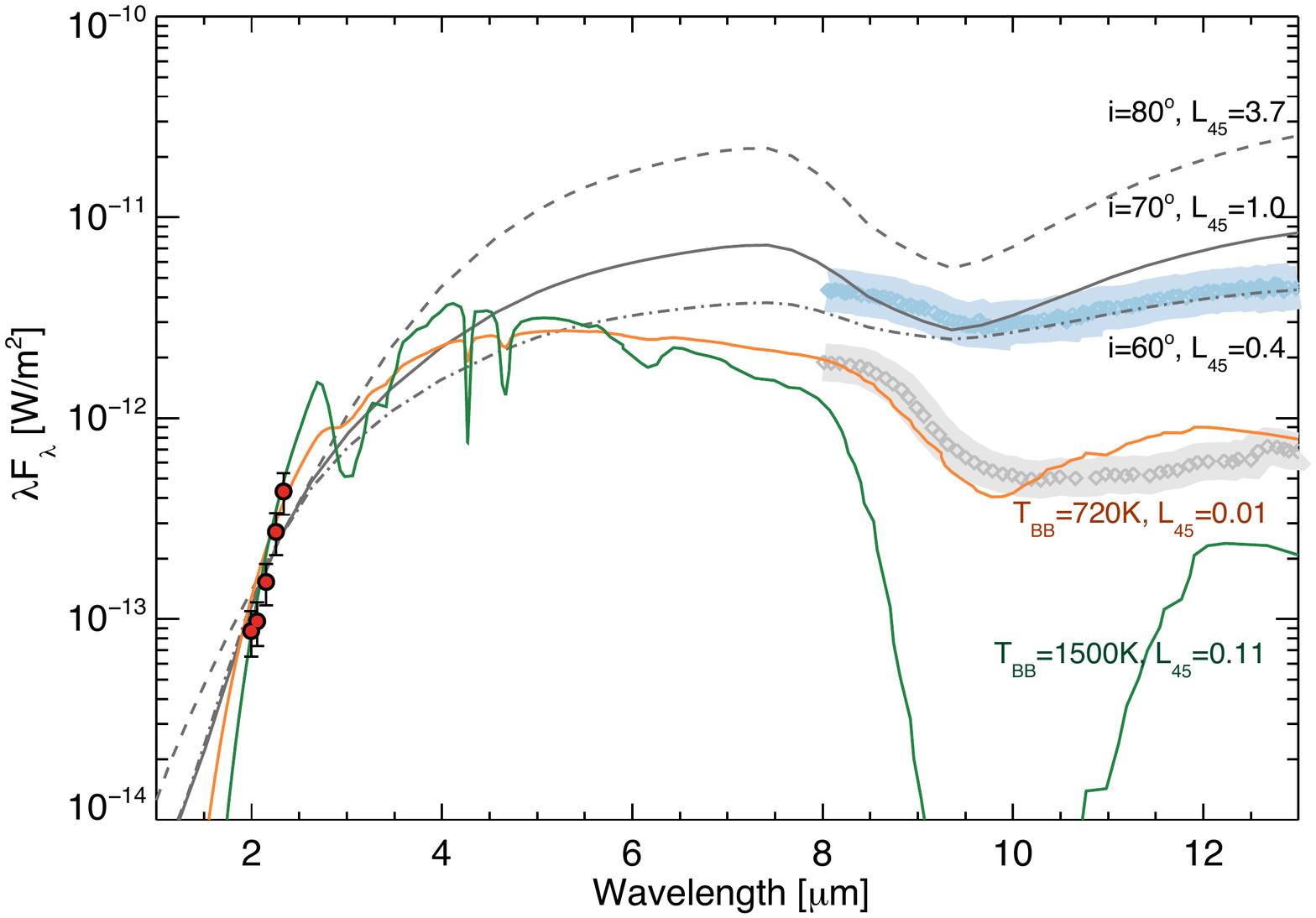}
\vspace{-2.5cm}
\caption{The K-band flux density observed with GRAVITY is shown in filled circles. The MIR flux density (incl. uncertainty) taken from \cite{raban2009} is shown in grey (central 20\,mas component only) and in blue (total MIR flux observed with MIDI within few 10\,pc). In black, we show three torus models from \cite{stalevski2012} with different inclinations of $i=60^\circ$ (dash-dotted), $i=70^\circ$ (solid) and $i=80^\circ$ (dashed). The two single temperature black body models are overplotted. Model 2 in orange ($T_{\rm eff}=720$\,K , $r=0.24$\,pc and $dr=0.045$\,pc.). Model 3 in green ($T_{\rm eff}=1500$\,K, $A_K=5.5$, $r=0.24$\,pc and $dr=0.026$\,pc). The bolometric luminosities of the different models are indicated in units of $\rm L_{45}=L_{bol}/10^{45}\,erg/s$.}
\label{fig:Fig5}
\end{figure*}

Our K-band image of NGC~1068 shows the central emitting region at an unprecedented spatial resolution. This allows us to estimate the brightness temperature based on the observed NIR and MIR fluxes. The GRAVITY fringe-tracker samples the K-band with five spectral elements between 1.95 and 2.35\,$\rm \mu m$. The observed flux of NGC~1068 shows a very red colour across the K-band. Based on the observed slope we can constrain the effective temperature and the extinction. Figure\,\ref{fig:Fig5} shows the observed K-band flux. We also included the published MIR (8-13\,$\rm \mu m$) SED for the central component (taken from \citealt{raban2009,lopez-gonzaga2014}). 

In the following subsections, we discuss several models that might explain the observed K-band emission. We first provide a short description of the key points of each model, where it can succeed, and where it fails. We focus on whether each of the models can explain the ring-like structure seen in the interferometric data, whether they match the slope of the NIR continuum, and whether they are consistent with other observational constraints (e.g. extinction, luminosity and covering factor).
We conclude on which model we favour in Sec.~\ref{sec:discussion}, where we provide a heuristic view of the central regions.

Model~1 is a geometrically thick clumpy torus composed of dusty clouds. This class of models is inconsistent with several aspects of the GRAVITY data. Such models, when viewed as close to edge-on as the orientation in NGC~1068 requires, cannot reproduce a ring-like structure. In addition the NIR continuum slope is rather shallower than that observed.

Model~2 enables us to explain both the observed K-band and MIR emission with a single, rather cool dust structure of $T_{\rm eff}\approx720$\,K. The required NIR luminosity is rather low, and hence so is the implied covering factor, which is naturally the case for a co-planar thin disc. The major difficulty is that this also implies too little foreground extinction.

Model~3 assumes hot dust of $T_{\rm eff}\approx1500$\,K absorbed by significant foreground extinction. This is mainly motivated by the observed emission size, which suggests hot dust close to the sublimation region. 
This model requires that the bulk of the MIR continuum originates in separate structures (consistent with the model proposed by \citealt{vollmer2018}). 
Its positive aspects are that it can match the observed steep red slope of the NIR continuum, and that the required foreground extinction is consistent with that expected from other observations. 
The main difficulty is that the high NIR luminosity requires a large covering factor. But this can be resolved if the accretion disc is tilted with respect to the maser disc as hinted at in point (4) of the introduction.

Model~4 is an alternative to Model~3, in which all of the bright NIR continuum observed by GRAVITY originates from the far side, rather than near-side, of the inner disc. Rather than recognising the K-band ring as a physical structure as in Model~3, this model assigns the NIR morphology to chance alignment due to clouds along the line of sight.
In particular, it requires unrelated patchy obscuration to create an apparent ring-like structure with a position angle, aspect ratio, and size that match the geometry given by the maser disc and dust sublimation radius.

\subsection{Model 1: geometrically thick clumpy torus}

We have already seen in Section~\ref{sec:data} that the ring-like structure, especially with the near side brighter than the far side, is hard to reconcile with a geometrically thick clumpy torus model that is close to edge-on. 
Similarly, these models require a careful matching of the vertical density structure to remain consistent with the presence of the thin maser disc at a radius $\la 1$\,pc.

Here we consider the photometric aspect. We compare our infrared SED to the torus models of \cite{stalevski2012}. These models use a two-phase medium of high density clumps and low-density gas filling the space in between. The model assumes standard silicate and graphite ISM-type grains and does not include a pure graphite grain component at small distances. The models have been calculated for a primary source of $10^{11}\,L_\odot$ ($L_{\rm bol}=0.4\times 10^{45}$\,erg/s) with inner and outer radii of 0.5 and 15\,pc, respectively. We scaled the models to the distance of NGC~1068 and to the observed K-band flux. In Figure\,\ref{fig:Fig5}, models with three different inclinations are compared to the observed SED. We find the best matching models to be at inclinations between $60^\circ$ and $70^\circ$. Models with higher inclinations (e.g. $80^\circ$) tend to over-predict the MIR emission. Models with lower inclinations, on the other hand, under-predict the MIR emission. These scaled models would imply a bolometric luminosity between $L_{\rm bol}=0.4-1.0\times 10^{45}$\,erg/s for $i=60^\circ-70^\circ$, in good agreement with earlier estimates of the bolometric luminosity (see Sec.\,\ref{sec:bol_lum}).
However, they do not match the red slope of the NIR continuum (with the caveat that the model is missing a pure graphite-grain component).

This model has difficulties reproducing the ring-like dust structure, accounting for the maser disc, and a NIR continuum slope that is too shallow.
We therefore rule out a geometrically thick clumpy torus model for NGC~1068.

\subsection{Model 2: cool dust disc}

MIR interferometry \citep{jaffe2004,raban2009,lopez-gonzaga2014} was used to model the central component with $600-800$\,K dust. In the cool dust model, we try to explain the observed K-band and MIR emission with a single source and temperature. The observed SED can be well approximated by a black body with an effective temperature of $T_{\rm eff} =720\pm30$\,K seen through a screen of extinction of $A_{\rm K}\approx0.9\pm0.3$ (see Figure\,\ref{fig:Fig5}). It is somewhat surprising that it is possible to explain the $2-13\,\rm \mu m$ SED with a single absorbed black body. In particular, the strong silicate absorption dip at $10\,\rm \mu m$ depends sensitively on the amount of dust and the dust composition. The very good agreement of the model and the observed silicate dip would indicate that the dust composition in NGC~1068 is very similar to the dust in our Galactic Centre. 

The minimum emission surface would be at least $\sim \rm 0.068\,pc^2$, which would correspond to a ring with $r=0.24$\,pc and a thickness of $\delta r=0.045$\,pc (0.65\,mas). Such a large thickness would be partially resolved and is therefore barely consistent with the data. A single black body ring would produce 1 percent of the estimated bolometric luminosity ($L=\sigma A T^4 = 1 \times 10^{43}$\,erg/s).
This covering factor is consistent with a thin dusty disc having $h/r < 0.14$, that is co-planar with the accretion disc even under the assumption of anisotropic emission.

The cool dust model would allow us to explain the near-to-mid infrared SED with a surprisingly simple model. However it predicts too little extinction, which is inconsistent with the $\rm Br\alpha$ constraint from ISO ($A_{\rm Br \alpha}>2.4$, \citealt{lutz2000a}) as well as the millimetre constraints on the column density \citep{tacconi1994, sternberg1994,garcia-burillo2017,imanishi2018}. The low extinction value results in a remarkably low bolometric luminosity of the dust. Moreover, it is hard to explain a lack of a dust emission component with a much higher temperature given the distance of the central ring from the very luminous nucleus. 
We therefore disfavour the cool dust and low extinction model.

\subsection{Model 3: hot dust disc}

The size of the observed structure in Figure\,\ref{fig:Fig3} suggests a temperature close to the dust sublimation temperature of $T_{\rm eff}=1500$\,K to 1800\,K. In order to reproduce the K-band slope, we therefore assumed a black body of $T=1500$\,K, which is highly absorbed (by foreground dust) with a mean K-band extinction of $A_{\rm K}=5.5\pm1.7$. Figure\,\ref{fig:Fig5} shows the observed fluxes and the proposed model. The corresponding optical extinction is $A_{\rm V}\approx90\pm30$. 
This is fully consistent with the extinction to the BLR derived from limits on the Br$\alpha$ broad line emission by \cite{lutz2000a}, as well as other estimates of the column density towards the nucleus noted in Sec.~\ref{sec:column}.

In order to explain the observed K-band flux with the hot dust model, the emission surface needs to be at least $\sim \rm 0.042\,pc^2$. If the emission is not a black body but a grey body then the emission region can be correspondingly larger. The minimum required surface is equivalent to a ring with $r=0.24$\,pc and a width of $\delta r=0.026$\,pc (0.38\,mas). This means that the observed ring-like structure is sufficient to explain the observed K-band flux. The width of the ring would be unresolved, which is also consistent with the observations. 

The hot dust model fails to produce the observed MIR emission. This is, in principle, not a problem, but it means that the bulk of the 8-13\,$\rm\mu m$ emission must originate from significantly cooler dust at larger radii, which is directly exposed to the central continuum and could be related to the dusty outflow \citep{cameron1993,bock2000,vollmer2018}.

The model also requires about 10 percent of the estimated AGN bolometric luminosity to be reprocessed by dust and re-emitted at NIR wavelengths ($L=\sigma A T^4 = 1.1 \times 10^{44}$\,erg/s).
This would be inconsistent with a system in which the accretion disc and thin dusty maser disc are co-planar due to the small covering factor.
However, there are clear indications that these structures are misaligned: the relative position angle of the radio jet and maser disc (see Figure\,8 of \cite{gallimore2004} for an illustration), as well as the inclination of the maser disc with respect to the orientation of the outflowing clouds \citep{kishimoto1999}.
Adopting anisotropic emission from the accretion disc as described by \cite{netzer2015} $L\propto {\rm cos}(\theta)[1+b~ {\rm cos}(\theta)]$ with $b\simeq2$, we find that about 10\% of the AGN luminosity is intersected by a surrounding disc with $h/r < 0.14$ (see Section~\ref{sec:discheight}) when the accretion disc is tilted by $40^\circ$. This tilt is consistent with position angle and inclination discrepancies above. A weaker anisotropy, for example $L \propto {\rm cos}(\theta)$, requires a somewhat smaller tilt.
This means that the `covering factor' issue, which is often taken to imply a geometrically thick structure can, for the NIR continuum, be entirely accounted for by a warped inner disc at scales $\la 0.1$\,pc.

In the scenario of a thin disc screened by foreground extinction, the visibility of the `central engine' (accretion disc and BLR) inside the sublimation hole is an obvious question. The K-band contribution of the accretion disc can be estimated based on the bolometric luminosity $L_{\rm bol}\approx1\times10^{45}$\,erg/s, the distance 14.4\,Mpc, a $\lambda5100\AA$ bolometric correction factor $\sim$9 \citep{netzer2015} and a SED slope of the accretion disc of $F_\nu\propto\nu^{\alpha}$ with $\alpha=-0.44$ \citep{vandenberk2001}. Also considering the A$_K \sim 5.5$ foreground extinction, we infer a K-band accretion disc flux of $\sim$9\,mJy. The empirical slope of \cite{vandenberk2001} is significantly redder than the canonical \cite{shakura1973} slope of $\alpha=0.33$. The K-band accretion disc emission for the canonical slope would be much weaker, and so the inferred contribution is probably a conservative upper limit. Other evidence for yet bluer slopes as found by \cite{kishimoto2007,kishimoto2008} would further enhance that argument.
The estimated accretion disc contribution of $\sim$9\,mJy corresponds to 6 percent or less of the flux in the reconstructed image (Fig.\,\ref{fig:Fig4}). This means that a point-like accretion disc could easily be hidden in the image including the centre of the dust ring. 

Both aspects of (1) accretion disc and (2) BLR visibility are consistent with the hot dust scenario. However, in both cases expected signatures are not far from the observational results/limits.   

\subsection{Model 4: pptically thick hot dusty disc} \label{sec:model4}

Model~3 above places the AGN at the centre of the K-band hot dust ring that has the expected position angle and aspect ratio for the inner geometry of NGC\,1068 and a size comparable to the dust sublimation radius, and identifies this position with the maser kinematic centre.
It implies the NIR continuum comes from both sides of the central disc, with the brightest on the near side while the far side is fainter.
This implicitly requires that the re-radiating clouds are optically thin to the hot dust emission.
Applying a standard gas-to-dust ratio as described in Sec.~\ref{sec:column} and the extinction law from Sec.~\ref{sec:extinction} mean that $\tau_{2.2\mu m} = 1$ occurs at a column density of $4\times10^{22}$\,cm$^{-2}$.
This is the maximum column density allowed for a typical cloud in the dust disc for those models.
In order for the far side of the disc to be fainter, there must be some additional obscuration to that line of sight.
The modest factor 4 difference in intensity is plausibly equivalent to 1--2 such clouds.
Confirming this alignment requires precise astrometry, which is addressed in Sec.~\ref{sec:astrometry}.

\begin{figure*}[ht]
\vspace{0cm}
\includegraphics[width=\textwidth]{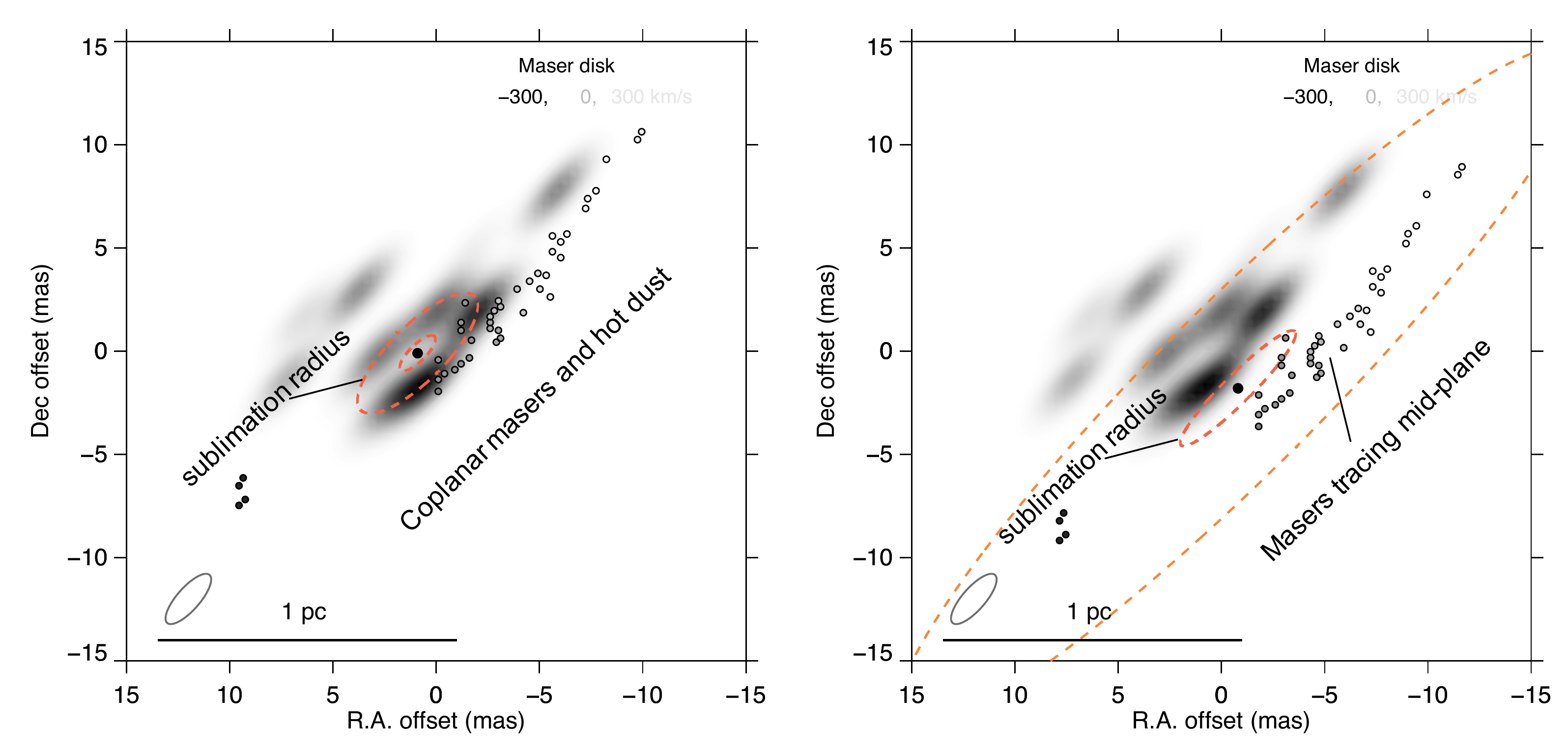}
\vspace{0cm}
\caption{Comparison of the alignment of the hot dusty disc models. The models themselves are not shown, only the implied interpretation of the observed continuum features. Left: Model~3 is the favoured interpretation, in which the ring-like structure in the K-band continuum arises from (partially) optically thin clouds (same as in Fig.\ref{fig:Fig4}). In this case, it traces a disc that has a similar position angle and aspect ratio to the inner geometry of NGC\,1068, is co-planar with the maser emission and traces the dust sublimation ring. Right: Model~4 comprises optically thick clouds, and the alignment of the mid-plane (indicated by the large dashed ring) to the near-infrared continuum observed with GRAVITY is given by radiative transfer simulations using the CAT3D-WIND model \citep{hoenig2017}.
In Model~4, the near side of the mid-plane as traced by the masers, but is completely obscured in the K-band. The K-band emission originates from the $\tau=1$ surface of the hot disc, that is still geometrically thin. But all the structure in the image is assigned to chance effects of patchy obscuration along the line of sight.}
\label{fig:Fig6}
\end{figure*}

An alternative alignment would be implied if the column density through individual clouds were higher, $N_H = 10^{23}$\,cm$^{-2}$ or more, as inferred from physical considerations of cloud stability in the vicinity of an AGN \citep{krolik1988,vollmer2004,hoenigbeckert2007}, so that they are optically thick at NIR wavelengths.
In this scenario, the NIR radiation is anisotropic and primarily emerges from the hot, AGN-facing side of each cloud. As a consequence, a ring consisting of such clouds would be brighter on the far side than the near side. We would be exposed to the directly heated surface of clouds on the far side, while on the near side we would mostly observe the clouds' cool back sides.
The hot dust disc is still co-incident with the maser dust, but these phenomena trace different phases of the dusty gas within the disc, so that the alignment shown in Fig.~\ref{fig:Fig4} is lost.
Models for NGC\,1068 constructed in this way could be similar to the disc-wind scenario proposed by \cite{hoenig2017} for NGC\,3783, which, in contrast to conventional geometrically thick torus models, indicate that the bulk of the MIR continuum arises in a wind.
Such a model would posit that the NIR continuum comes from the $\tau_{2.2\mu m} = 1$ surface of the hot disc that is still geometrically thin, although offset from the equatorial plane according to $h/r$, which could be in the range 0.1-0.2; and that at longer wavelengths pertinent to the masers, the disc is optically thin.
In this model, the observed NIR continuum would still extend inwards to the dust sublimation radius, but the brightest spot observed in the K-band by GRAVITY would coincide with the far side of the disc.
The ring morphology with radius matching the expected dust sublimation radius, the similarity between the maser disc and the K-band emission, as well as the alignment of the maser and radio continuum with the K-band would all be chance associations. 
The right panel of Figure~\ref{fig:Fig6} shows the alignment for the optically thick hot dust disc model.
The alignment of the GRAVITY data to the disc mid-plane has been implied by radiative transfer simulations using the CAT3D-WIND model \citep{hoenig2017} with an assumed inclination of $i=80\deg$, a scale height of the disc of $h/r=0.1$, and the optical depth of an individual cloud of $\tau_V=50$.

For this model, the resulting structures originate with the random distribution of clouds, and hence are due to variations in obscuration along the line of sight. As such, each instance of this model that is created will look different. This makes the model seem arbitrary because one needs to create many such instances in order to pick one that approximates the characteristics of the observed continuum distribution.

\section{Discussion}
\label{sec:discussion}

\subsection{Astrometry}
\label{sec:astrometry}

The required accuracy to make a decisive statement on the astrometry of the K-band continuum with respect to the masers or radio continuum is on the order $\leq$ 2-3\,mas. Without reference features, which can be identified simultaneously in the K-band and in the radio, absolute astrometry is required to relate the radio and infrared measurements. The radio position is sufficiently well known with an uncertainty $\sigma<1.5\,$mas \citep{gallimore2004}. Unfortunately the astrometric uncertainty of the infrared data is a factor 10 worse ($\sigma\sim20\,$mas). Future interferometric observations should be able to exploit nearby stars from the GAIA catalogue as astrometric references. Then, by constructing a local baseline solution using these reference stars and by logging the optical path difference between the stars and NGC~1068, a sufficiently accurate astrometric solution might be achieved.

Without such an astrometric reference yet available, we favour Model~3 which provides a natural explanation for several aspects of the observed ring-like structure:(i) the match of its position angle and aspect ratio to those of the maser disc; (ii) the match of its radius to the dust sublimation region around the AGN; and (iii) the alignment between its near-side and the masers, when the centre of the ring is aligned to the black hole location derived from the maser kinematics (see Sec.~\ref{sec:maser}).
The hot dust disc, with clouds that are optically thin in the NIR, is consistent with observations and provides a rationale for the similarity of the masers, radio continuum and the hot dust K-band emission. Based on that model we propose a simple disc-like geometry, which is obscured by foreground (within a few parsec) molecular gas.

\subsection{A heuristic model for the central region of NGC 1068}

\begin{figure*}[ht]
\vspace{0cm}
\centering
\includegraphics[width=12cm]{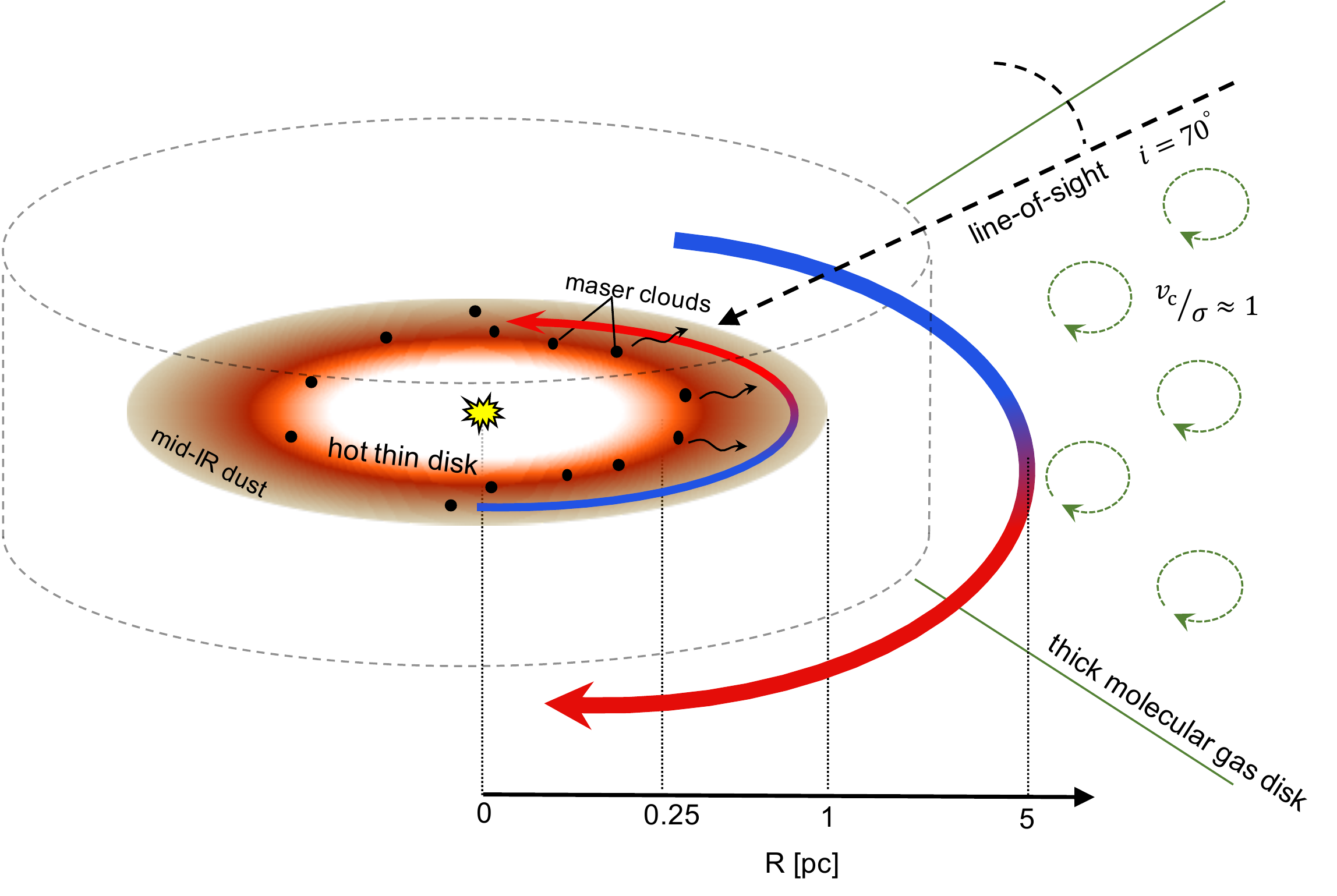}
\vspace{0cm}
\caption{Sketch of the observed central structures. The K-band emission traces the inner rim of a thin disc of hot gas and dust, at or close to the dust sublimation radius $r\approx 0.24\,\rm pc$. The inner water masers are cospatial with the hot K-band dust. The masers stretch out to $r\approx1\,\rm pc$ \citep{gallimore2001}.  MIR observations show warm dust on similar scales as the outer masers \citep[e.g.][]{raban2009}, likely originating from the disc periphery. ALMA observations of HCN and HCO$^+$ show a turbulent structure, which rotates in the opposite direction as the maser disc \citep{imanishi2018}. The $v_c/\sigma$ of the molecular gas structure argues for a thick disc, which contains enough gas mass to reach column densities of $N_H\approx10^{23}\,\rm cm^{-2}$ that screen the central region by $A_V\approx90$ ($A_K\approx5.5$) from the observer. }
\label{fig:Fig7}
\end{figure*}

With a wealth of observations across the electromagnetic spectrum at hand, we try to construct a simple heuristic picture of the central few parsecs of NGC~1068 (see Fig.\,\ref{fig:Fig7}).
The K-band emission traces the hot dust sublimation radius of a thin disc, seen at an inclination of 70$^\circ$. The inner masers are cospatial with the K-band emission. The masers extend out to radii of $\sim$1\,pc \citep{gallimore2001}, similar to the size of the warm dust structure ($\sim$600\,K) seen at MIR wavelengths \citep{jaffe2004,raban2009,lopez-gonzaga2016}. This suggests that the warm dust originates from the outer region of the same thin disc. Beyond $\sim$1\,pc distance no masers have been observed, probably because the conditions disfavour masing (e.g. too low dust temperature, $<400$\,K). In contrast to the thin disc at scales of $<$1\,pc, ALMA observations by \cite{imanishi2018} reveal a significant amount of molecular gas, which is rotating with $v_c\approx 40-50$\,km\,s$^{-1}$ (notably in the opposite sense as the maser disc) at 5--10\,pc distance. The comparably large velocity dispersion of $\sigma\approx30-40$\,km\,s$^{-1}$ and the correspondingly high $\sigma/v_c\simeq1$ suggests a thick disc with a large scale height of $h_z/R\simeq1$. The disc contains enough gas mass to reach column densities of $N_H\approx10^{23..24}\,\rm cm^{-2}$ that screen the central region by $A_V\approx90$ ($A_K\approx5.5\pm1.7$). The central few parsecs are fed by two gas streamers \citep{muller-sanchez2009}, which come very close to the nucleus. The northern and southern streamers reach pericentre distances of $\approx$5\,pc and $\approx$\,1pc repectively. Their opposite sense of rotation could contribute to the turbulence observed in the molecular disc.
  
\section{Conclusion}

We present new 0.2\,pc resolution observations from the GRAVITY interferometer on the VLT, which spatially resolve the hot dust continuum in the central parsec of NGC~1068. The following conclusions assume a distance of 14.4\,Mpc. We find that:

\begin{itemize}

\item
The dust structure is dominated by a thin, ring-like structure with a radius of $r=0.24\pm0.03$\,pc, a scale height $h/r<0.14$, an inclination $i=70^\circ \pm5^\circ$ and a position angle of $PA=-50^\circ \pm 4^\circ$.

\item
The spatial scale of the ring agrees with the predicted sublimation distance of graphite dust illuminated by a $L_{\rm bol}\approx 0.4-4.7\times10^{45}$\,erg/s AGN (radiating at or in excess of its Eddington limit). We therefore associate this with the dust sublimation region.

\item 
The near-infrared continuum structures show a striking resemblance to the maser disc. We suggest that both types of emission originate from the same region at $\sim$ 0.4\,pc distance from the central source. The maser disc is co-spatial with the south-western rim of the dust sublimation region, the side closer to the observer (as expected due to the maser beaming).

\item
Based on the observed ring-like structure and steep near-infrared continuum slope, we disfavour geometrically thick clumpy torus models (which also do not account for the maser disc).

\item
The dust structure and photometry are consistent with a simple model of hot dust at $T\sim1500\,\rm K$ that is behind $A_K\sim 5.5$ ($A_V\approx90$) mag of foreground extinction. This amount of screen extinction is expected from an upper limit to the $Br\alpha$ broad line and could be provided by the dense and turbulent gas distribution observed on scales of 1-10\,pc.
The model implies that the near-infrared and MIR emitting dust are in spatially separate locations (the maser disc and the outflow respectively).

\end{itemize}


\begin{acknowledgements}
We thank the referees for their careful reading of the manuscript and their suggestions that have helped to improve it.
This research has made use of the NASA/IPAC Extragalactic Database (NED), which is operated by the Jet Propulsion Laboratory, California Institute of Technology, under contract with the National Aeronautics and Space Administration. This work has made extensive use of the software {\it QFitsView}, which has been developed by Thomas Ott, Max Planck Institute for Extraterrestrial Physics (\url{http://www.mpe.mpg.de/\~ott/QFitsView/}). 
F.E. and O.P. acknowledge support from ERC synergy grant No. 610058 (Black-HoleCam). J.D. was supported by a Sofja Kovalevskaja award from the Alexander von Humboldt foundation and in part by NSF grant AST 1909711. A.A. and P.G. acknowledge funding from Funda\c{c}\~ao para a Ci\^encia e a Tecnologia through grants UID/FIS/00099/2013 
and SFRH/BSAB/142940/2018. P.O.P acknowledges support from the Programme National Hautes Energies (PNHE) of the Centre National de la Recherche Scientifique (CNRS).
\end{acknowledgements}

\bibliography{references}



\end{document}